# Heat transfer in a planer diverging channel with a slot jet inlet

Md Insiat Islam Rabby [1], Mohammad Ali Rob Sharif[2], Mohammad Tarequl Islam[1], Md. Rakidul Islam[1], Md Mahidul Alam[1]

[1]*Department of Mechanical Engineering, Military Institute of Science and Technology, Bangladesh*
[2]*Aerospace Engineering and Mechanics Department, The University of Alabama, USA*

*Corresponding author: insiatislam8@gmail.com

**Abstract.** This article delves into a numerical exploration of two-dimensional, incompressible, laminar flow within a confined diverging jet. The study aims to understand how variations in the inlet opening fraction and Reynolds number affect the heat transfer and flow patterns. The research employs the finite volume method with a collocated mesh to solve the governing equations. Across a broad spectrum of inlet opening fractions (0.2, 0.4, and 0.6) and Reynolds numbers (ranging from 500 to 900), the findings reveal that increasing the inlet opening fraction of the jet in the diverging channel can lead to a remarkable (53%) improvement in heat transfer while simultaneously reducing pressure loss by 90%. This outcome holds the potential to conserve energy by requiring less pumping power. Notably, this investigation is pioneering and offers novel and valuable insights into enhancing heat transfer through the utilization of a diverging channel.

## INTRODUCTION

A broad range of engineering applications, including electronics, solar collectors, and internal combustion engines, generate heat, which can either be advantageous or detrimental depending on the specific application. In the realm of thermal engineering, various equipment, such as heat exchangers, are employed for both cooling and heating processes. However, these equipment face limitations in terms of efficiency due to inherent factors like heat dissipation characteristics and operational modes. Consequently, there is a pressing need for innovative ideas, techniques, and designs in recent decades, driven by the global surge in energy consumption, to effectively dissipate heat, maintain optimal operating temperatures, and enhance overall efficiency.

In recent decades, numerous researchers have delved into the subject of convective heat transfer within conduits featuring varying cross-sectional areas, as evidenced by studies [1–12]. Yang and Price [2] conducted a numerical investigation on laminar flow and heat transfer in the entrance region of a converging plane-walled channel, considering both angular and radial directions, while maintaining a uniform wall temperature. They examined three different taper angles and observed that as this angle increased, the heat transfer rate increased, and the pressure drop decreased due to the expansion in the channel's surface area.

Su and Lin [6] explored the heat transfer and pressure drop characteristics in convergent and divergent ducts with a rectangular cross-section, keeping the wall temperature constant. Their numerical findings revealed an anticipated trend: an increase in the convergence angle led to higher Nusselt numbers and pressure drop, while an increase in the divergence angle resulted in reduced pressure drop and heat dissipation.

Liu and Gau [9] undertook experimental research into the initiation and progression of buoyancy-driven secondary airflows and the enhancement of mixed heat transfer in horizontal convergent and divergent channels. They discovered that, unlike the divergent channel, the convergent channel experienced a localized decrease in mixed convection, resulting in smaller and more stable plumes. Although this may yield a somewhat smaller improvement in heat transfer, it was observed that average Nusselt numbers could be increased for both decelerating and accelerating flows, especially at higher Reynolds numbers.

As far as the authors are aware, there is a lack of existing research in publicly available sources concerning fluid flow and heat transfer in divergent channels using jet flow. In this current study, our primary emphasis will be on assessing how the inlet fraction and Reynolds number influence the flow and heat transfer within these channels. The aim of the present study is to explore potential enhancements that can be attained and identify any constraints that may limit such improvements.

## MODEL DESCRIPTION AND BOUNDARY CONDITIONS

The two-dimensional flow configuration in the *xy* plane is shown in Figure 1. It consists of a diverging channel of length l where the principal flow direction is from left to right. The diverging channel walls have a uniform heat flux of 2500 W/m². The inlet slot jet of width, *d*, is located symmetrically at middle of the left boundary, while the outlet of same width *d* is located symmetrically at the middle of the right boundary. The left and right boundaries are adiabatic excluding the inlet and outlet openings. The inlet jet width *d* is taken as the reference length for the problem. The width of the left boundary, *H*, is taken as 10 mm while the width the right boundary, *w*, is taken as 20 mm, and the channel length, *l*, is taken as 150 mm. The no-slip conditions are imposed on all walls. An important parameter for the problem is the inlet opening fraction, *IF = d/H*.

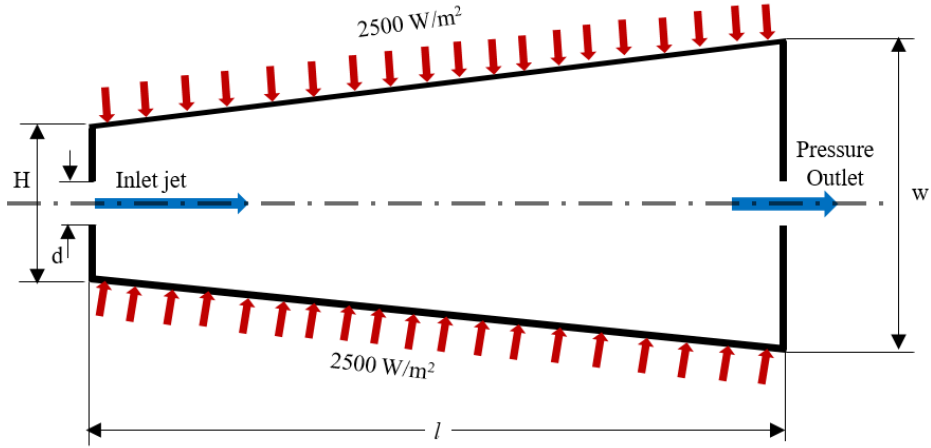

**FIGURE 1**. Problem configuration.

## GOVERNING EQUATIONS AND NUMERICAL SOLVING PROCEDURE

The governing equations for the two-dimensional laminar incompressible is given as:

Continuity Equation:
$$\nabla \cdot \vec{V} = 0 \qquad (1)$$

Momentum Equation:
$$\rho(\nabla \cdot \vec{V})\vec{V} = -\nabla P + \mu \nabla^2 \vec{V} \qquad (2)$$

Energy Equation:
$$\vec{V} \cdot \nabla T = \alpha \nabla^2 T \qquad (3)$$

where, $\vec{V}$ is the velocity vector, $T$ is the temperature, $P$ is the pressure, $\mu$ is the dynamic viscosity, $\rho$ is the density and $\alpha$ is the thermal diffusivity.

The flow Reynolds number is defined in terms of the inlet velocity $U_{in}$ and jet opening $d$ as Re = $\rho U_{in} d/\mu$ where the denominator corresponds to the mass flow rate per unit depth perpendicular to the flow plane. The mass flow rate is varied to achieve the desired Reynolds number form which the inlet velocity is obtained for a particular *d*.

A finite volume technique on a collocated grid is implemented for discretizing the governing equations inside the computational domain. The SIMPLE algorithm is used to link the pressure and velocity fields. For the stability of the solution, the diffusion term in the momentum equations is approximated by the second order central difference scheme while the convection terms are approximated by the second order upwind scheme. The governing equations are solved iteratively and sequentially until the normalized residuals for each equation falls below $10^{-6}$.

## GRID INDEPENDENCY TEST AND VALIDATION

Prior to the current analysis, a grid sensitivity test was performed to find a reasonable compromise between computational cost and accuracy. The mesh was refined successively until a mesh independent solution was obtained. Four different grid resolution were considered: with element numbers 43865, 66413, 84837, 111860, and 138351 as shown in Table 1. The relative difference in the average Nusselt number at the channel surface is only 0.79% when the grid was changed from 66413 to 84837. Therefore, the 84837 grid was chosen for the rest of the simulations.

**TABLE 1.** Grid independency test

| Element size | Surface Nusselt Number | Percent relative change |
|---|---|---|
| 43865 | 7.52 | - |
| 66413 | 7.62 | 1.330 |
| 84837 | 7.68 | 0.79 |
| 111860 | 7.76 | 1.04 |
| 138351 | 7.85 | 1.16 |

The computational validity was checked by comparing the results of the current solver with literature. The local Nusselt number distribution along the channel surface is compared with the experimental data for a channel between two parallel plates by Wen and Ding [13] for different Reynolds number as shown in Figure 2. The results of the current simulations are in close agreement to those available in the literature, thus indicating that both the solver and numerical scheme are accurate.

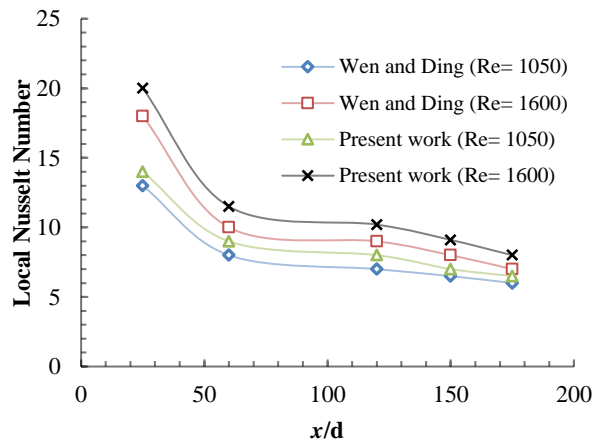

**Figure 2:** Comparison between present work and experimental work of Wen and Ding [13].

## RESULTS AND DISCUSSION

Simulations are conducted for a range of Reynolds numbers 500 to 900 and inlet fraction, IF = 0.2, 0.4 and 0.6 to determine the flow characteristics, Nusselt number, and pressure loss. Figure 3 shows the temperature contour in the flow domain for different Reynolds number. The figure illustrates maximum temperature around the top and bottom wall and minimum temperature at the middle of the channel for different Reynolds number. As constant heat flux applied at both top and bottom wall therefore maximum temperature was observed around the wall section which gradually decreases to middle section of the channel. By increasing the Reynolds number, the temperature differences from wall to inside fluid which drives the heat transfer rate for the system.

Furthermore, Figure 4 shows the flow streamlines at different Reynolds number for inlet fraction of 0.4. It is observed that the jet development is almost asymmetric. It is seen that asymmetric flow exhibits a wavy pattern with a larger wavelength and the flow development inside the duct occurs over a larger axial distance. Recirculation zone is also observed immediately after the inlet section and before the outlet section. The developed recirculation zone at the bottom section in outlet is the largest in shape and size which have potential impact in heat transfer argumentation.

Figure 5 and 6 show the streamlines and temperature contours for different inlet fraction at Reynolds number 500. Symmetric streamlines are observed for inlet fraction 0.6 while 0.2 and 0.4 inlet fraction showed asymmetric behavior. For inlet fraction 0.2 and 0.4 recirculation zones were observed at different sections of the channel while inlet fraction 0.6 did not present any recirculation zone. Inlet fraction 0.2 shows irregular recirculation zones. Additionally, from temperature contour it is observed that by increasing inlet fraction, temperature gradient is decreasing.

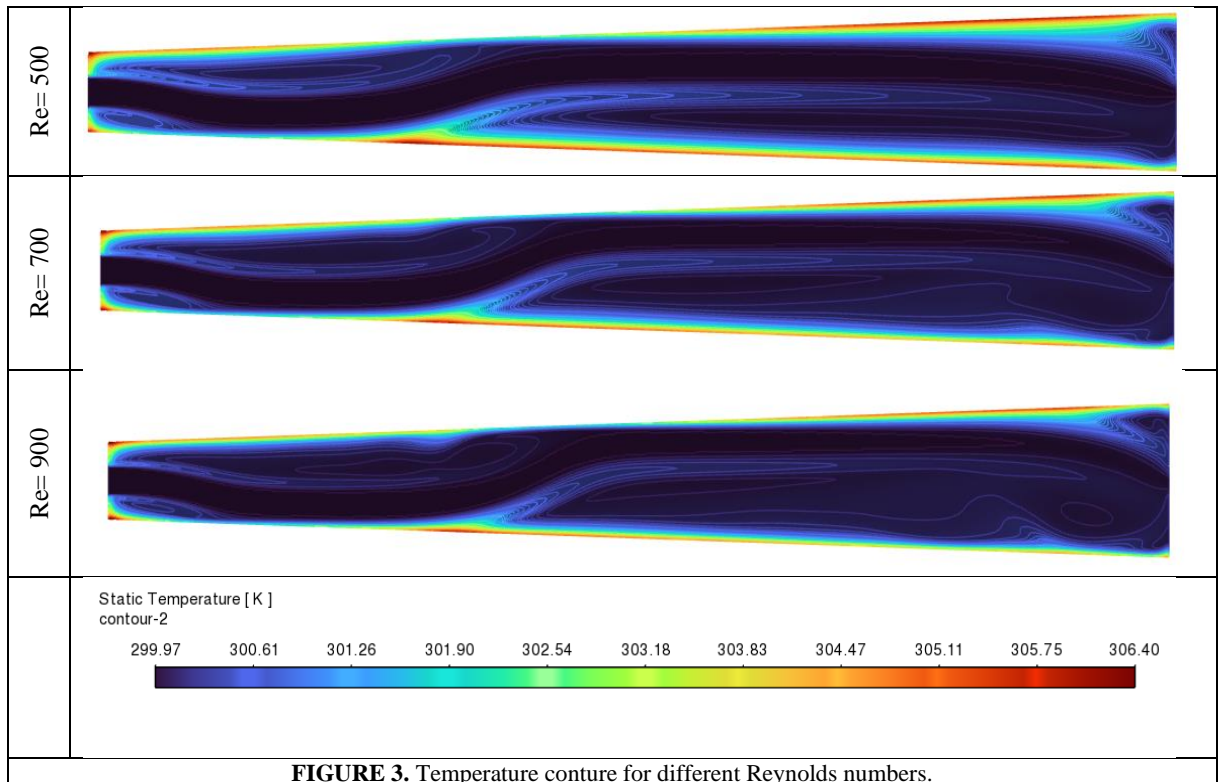

**FIGURE 3.** Temperature conture for different Reynolds numbers.

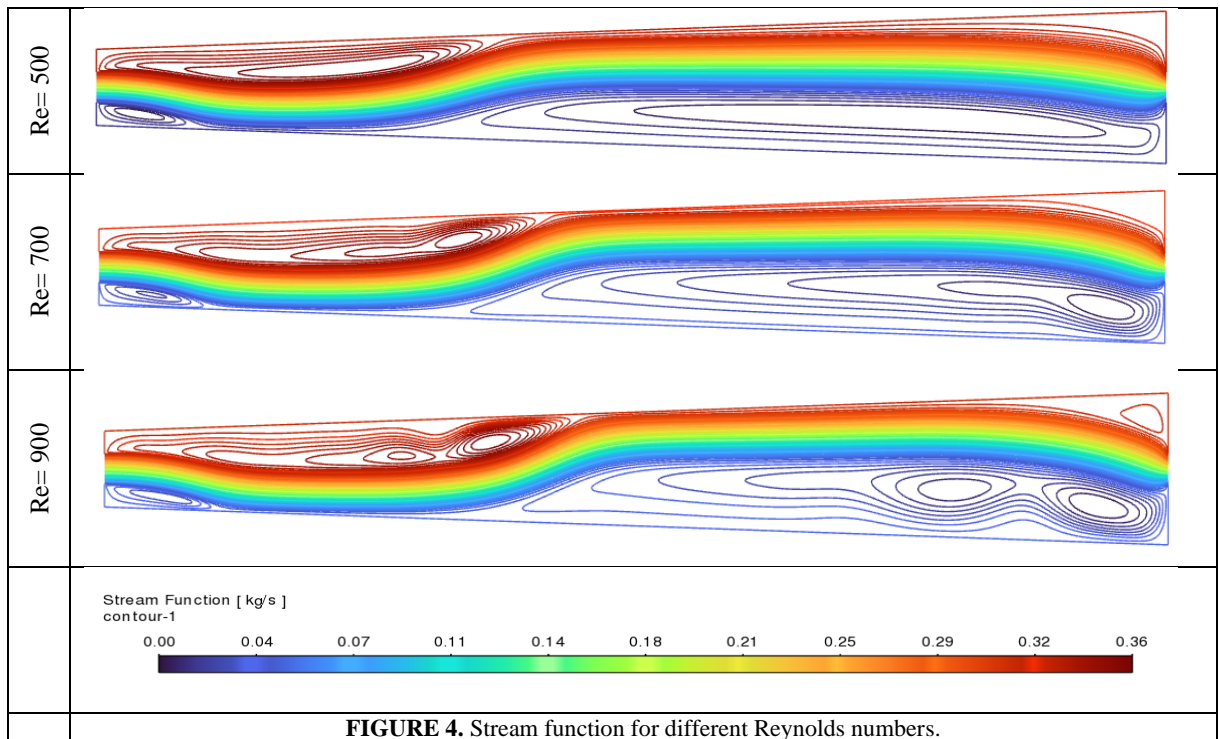

**FIGURE 4.** Stream function for different Reynolds numbers.

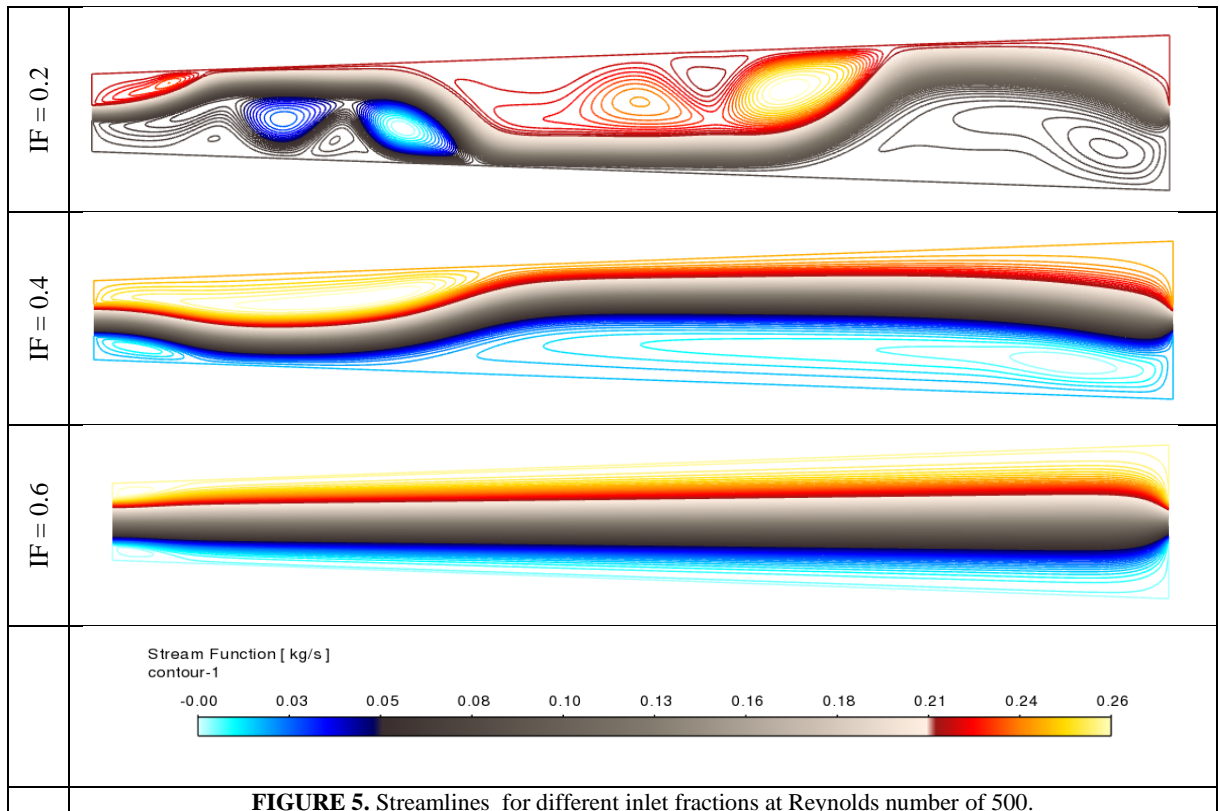

**FIGURE 5.** Streamlines for different inlet fractions at Reynolds number of 500.

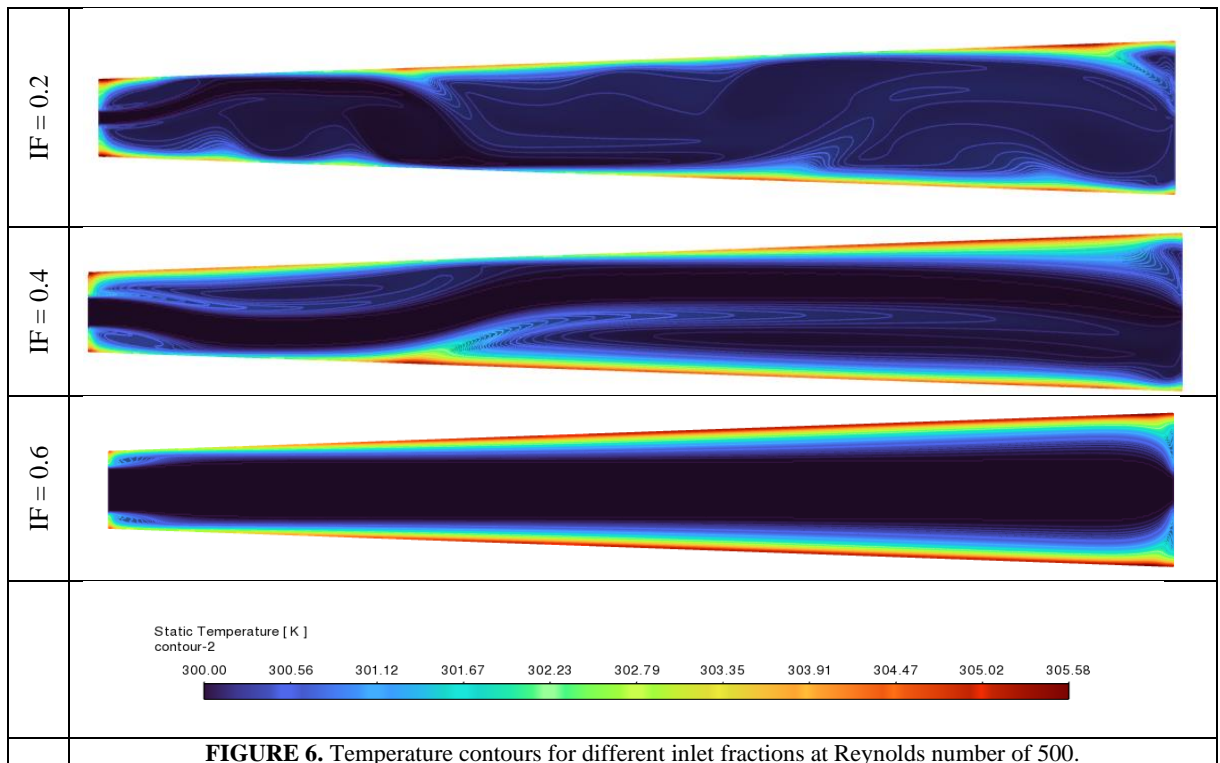

**FIGURE 6.** Temperature contours for different inlet fractions at Reynolds number of 500.

Figure 7 and 8 represent the average surface Nusselt number and pressure loss for different inlet fraction of jet flow and Reynolds number. With Reynolds number the surface Nusselt number and pressure loss increases gradually. By increasing the inlet fraction from 0.2 to 0.6 the Nusselt number enhances 30% to 53% for different Reynolds number while pressure loss decreases almost 90%. This finding refers that by increasing the inlet fraction of jet flow in diverging channel maximum 53% enhancement in heat transfer can be achieved with 90% reduction in pressure loss which will help in saving pumping power.

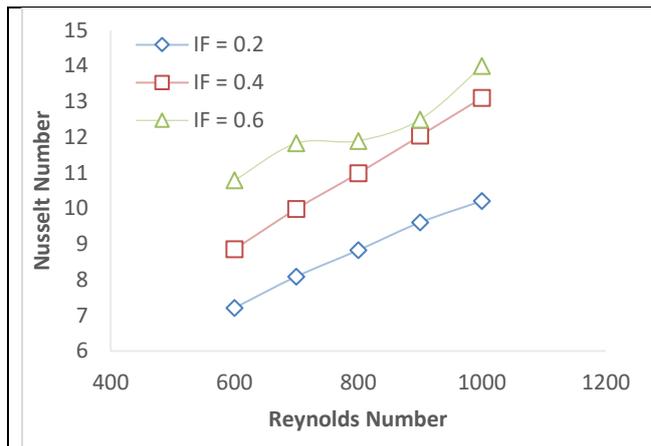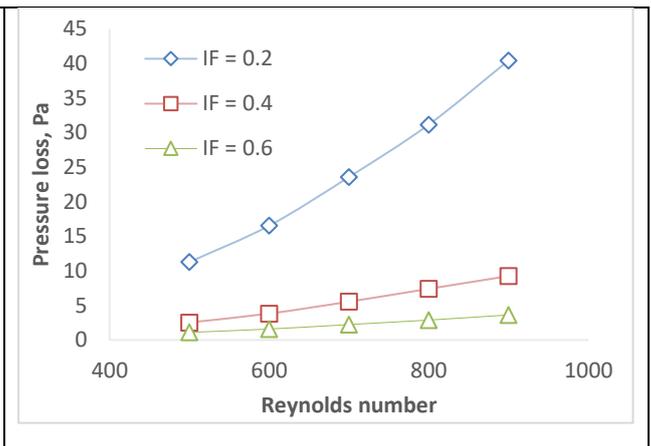

| FIGURE 7. Average Surface Nusselt number for different inlet fraction and Reynolds number. | FIGURE 8. Pressure loss for different inlet fraction and Reynolds number. |

## CONCLUSIONS

An investigation of heat transfer and fluid flow characteristics for a diverging channel by applying jet flow is presented. Effects of inlet opening fraction and Reynolds number have been addressed in the present study. The key findings are as follows:

1. Reynolds number and inlet fraction influences the stream function and temperature gradient inside the diverging channel by applying jet flow.
2. By increasing the Reynolds number Nusselt number and pressure loss also increases almost linearly.
3. By increasing the inlet fraction of jet flow in diverging channel maximum 53% enhancement in heat transfer can be achieved with 90% reduction in pressure loss which will help in saving pumping power.